\documentclass[aps,showpacs,epsfig,twocolumn]{revtex4}
\usepackage{epsfig}
\begin{document}

\title{\bf\Large {Spin-Fermion model of $UGe_2$}}
\author{Naoum Karchev\cite{byline}}

\affiliation{Department of Physics, University of Sofia, 1126 Sofia, Bulgaria}

\begin{abstract}

It is assumed that $U$ atoms in $UGe_2$ have a number of $f$ electrons appropriate to give
them each a spin $s=1$ as well as one extra itinerant electron which may equally well
be on one or other $U$ atom.
The dynamical degrees of freedom are spin-$s$ operators of localized
spins and spin-$1/2$ fermi operators of itinerant electrons.
Applying hydrostatic pressure changes the bandwidths of spin-up and
spin-down itinerant electrons in different way, which leads to decreasing of the contribution
of the fermions to the magnetization keeping the spin-fermion interaction unchanged. 
In turn the local spin-fermion interaction leads to ferromagnetic superconductivity. 
The model accounts, in a quantitative and natural way, 
for the characteristics of the coexistence of superconductivity and ferromagnetism in 
$UGe_2$, including many of the key experimental results: metamagnetic transitions, 
quantum transition from ferromagnetism to ferromagnetic superconductivity, the position 
of the highest superconducting critical temperature etc. 

\end{abstract}


\maketitle 

$UGe_2$ is the first example where ferromagnetism and superconductivity
coexist\cite{sfm1,sfm2}. The superconductivity is found experimentally only in 
ferromagnetic phase, and only in a limited pressure range($p'_c,p_c$). There are 
two successive quantum phase transition, from ferromagnetism to ferromagnetic 
superconductivity at $p'_c$, and at higher pressure $p_c$ to paramagnetism (fig1a).

As the pressure is increased there is an abrupt decrease of the ordered moment at $p_x$
($p'_c<p_x<p_c$) and another at $p_c$ (fig1b). The ferromagnetic state below $p_x$ is
referred to as FM2 and the high pressure ferromagnetic state as FM1\cite{sfm3}. It has
been suggested that a spin and charge density wave might be formed in the FM2 state, due
to the nesting of the Fermi surface, and they are responsible for the transition at 
$p_x$\cite{sfm4}. However, neutron diffraction studies have not detected any static
order due to a spin and charge density wave. Another possibility is that the transition
at $p_x$ is a result of a novel tuning of the Fermi surface topology by 
the magnetization\cite{sfm5}.

The temperature dependence of the magnetization in $UGe_2$ is quite different from that found
in weak itinerant ferromagnets. At zero pressure, above and well away from $p_x$ the low
temperature dependence of the magnetization has the form $M(T)/M(0)\sim [1-(T/T_c)^3]^{1/2}$
\cite{sfm6}. 
Strictly speaking, $UGe_2$ is not a weak itinerant ferromagnet, and the point
where ferromagnetism and superconductivity disappear simultaneously is 
not a quantum critical point at all. The Curie temperature $T_c$ decreases 
while the magnetization remains unchanged. For conventional
weak ferromagnets the Curie temperature scales with magnetization. 
$UGe_2$ differs mainly in having a stronger spin orbit interaction that leads
to an unusually large magneto-crystalline anisotropy with easy magnetization 
axis along shortest crystallographic axis. The differential susceptibility has been
measured, since it gives a measure of the spectrum of the magnetic excitations.
The main conclusion is that the differential susceptibility is strongly anisotropic in the 
high pressure FM1 and paramagnetic phases but weakly anisotropic in the low pressure
FM2 phase\cite{sfm6}.
It is plausible that increasing the pressure, one changes the anisotropy, which in turn
shifts the system from itinerant behaviour to a higher pressure phase which is dominated
by localized spins.

\begin{figure}[h]  
\vspace{0.03cm}  
\epsfxsize=5.8cm  
\hspace*{0.2cm}  
\epsfbox{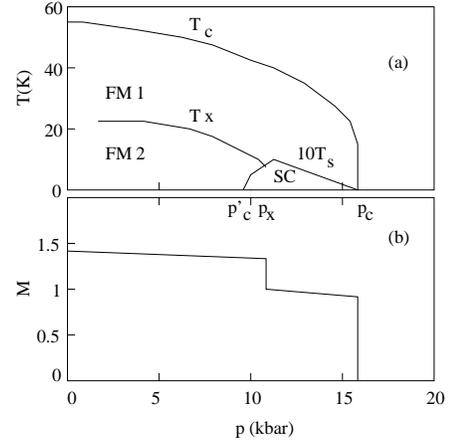}   
\caption{(a)The p-T phase diagram of $UGe_2$. $T_c$ is Curie temperature,
$T_s$ is the superconducting transition temperature.(b) Pressure dependence of the
dimensionless magnetization per lattice site.}  
\label{fig1}  
\end{figure}  

So far, there are no theoretical considerations of these complicated phenomena. 
The calculations have considered the superconductivity to appear from completely 
itinerant ferromagnetic state\cite{sfm7,sfm8}, or have been based on the physics of 
local moments\cite{sfm9,sfm10}.      

Motivated by the experimental findings, one assumes that $U$ atoms in $UGe_2$ 
have a number of $f$ electrons appropriate to give
them each a spin $s=1$ as well as one extra itinerant electron which may equally well
be on one or other $U$ atom.
The dynamical degrees of freedom are spin-$s$ spin operators $\textbf{S}_i$ of localized
spins  and spin-$1/2$ fermi operators $c_{i\sigma}$ of itinerant electrons, where
$i$ denotes the sites of a three dimensional lattice. 
The dimensionless magnetization $M=\mu/\mu_B$ of the system per lattice site at zero 
temperature 
is $M=s+m$ where $m$ is the 
contribution of mobile electrons. The parameter 
$m$ depends on the microscopic parameters of the theory and characterizes 
the vacuum. If, in the vacuum state, every lattice site is
occupied by one electron with spin up, then $m=1/2$, 
and the electrons are highly localized 
as in the uranium compounds known as "heavy-fermion systems".
When, in the vacuum state, some of the sites are 
doubly occupied or empty, then $m<1/2$ and the electrons are itinerant. 
The system approaches the internal 
point (IP) when $m\rightarrow 0\quad (M=s)$. 
It corresponds to the point $p_x$ of the phase diagram of $UGe_2$ (fig1).

The local spin-fermion interaction leads to an effective four-fermion interaction
 which in turn leads to p-type magnon-induced ferromagnetic superconductivity 
(FM-superconductivity)\cite{sfm8}. The order parameter is a spin anti-parallel 
component of a spin-1 triplet with zero spin projection 
($\uparrow\downarrow+\downarrow\uparrow$). The transverse spin 
fluctuations are pair forming and the longitudinal ones are pair 
breaking. The effective potential is attractive within an interval 
$(p_f-\Lambda,p_f+\Lambda)$, around the fermi surface $p_f$, 
where $\Lambda$ depends on the parameters of transverse and longitudinal
spin fluctuations. When the fermions contribute to the magnetization of the 
system ($m\neq0$) spin-up and spin-down electrons  have different  
(majority and minority) Fermi surfaces.  
If the Fermi momenta $p_f^{\uparrow}$ and 
$p_f^{\downarrow}$ lie within the interval $(p_f-\Lambda,p_f+\Lambda)$ the interaction 
between spin-up electrons, which contribute to the majority Fermi surface, and spin-down  
electrons, which contribute to the minority Fermi surface, is attractive. As a result,  
spin-up electrons from the majority Fermi surface transfer to the minority Fermi surface 
and form spin anti-parallel Cooper pairs, while spin-down electrons from the minority 
Fermi surface transfer to the majority one and form spin anti-parallel Cooper pairs too.
The domain between the Fermi surfaces determines the fermions' contribution to the 
magnetization $m$, 
but it is cut out from the domain of integration in the gap 
equation. When the electrons' contribution to the magnetization increases, the 
domain of integration in the gap equation decreases, and for some value of $m=m'$, 
respectively at magnetization $M'=s+m'$ the system undergoes a transition from 
FM-superconductivity to ferromagnetism ($p'_c$ point on fig1a). 
At IP ($m=0$) the domain of integration
in the gap equation is largest.
As a result the superconducting critical temperature 
is highest when the system is at IP.
 
The anisotropy modifies the spin-wave
excitations adding a gap in the magnon spectrum. Increasing the gap,
the pair formation as a result of magnons' exchange is 
suppressed, which in turn leads to decreasing of the superconducting critical 
temperature. Hence, the most appropriate assumption, which closely matches 
the experimental result, is that increasing 
the hydrostatic pressure one increases the magneto-crystalline anisotropy.  
For pressures above $p_x$ the contribution of the itinerant electron to 
the magnetization is zero ($m=0$), and
the magnetization is due to magnetization of the localized spins $M=s$. Hence, the
transition to the paramagnetism $M=0$ undergoes with jump only.  

The important point in the spin-fermion theory is the mechanism of driving the system
from a state with magnetization $M=s+m>s$ to the internal point (IP) ($M=s$). 
The subtle point is the spin-fermion interaction. It splits  
the spin-up and spin-down Fermi surfaces and leads to a nonzero contribution of itinerant
electrons to the magnetization. Driving the system to the internal point (IP)  one
has to compensate this overall shift in the relative position of the energy bands keeping 
the spin-fermion interaction unchanged.
In this paper a mechanism of compensation by means of different changes in bandwidths of
spin-up and spin-down electrons is considered. The Hamiltonian of the spin-fermion model is
\begin{eqnarray}
\hat H & = &
-J\sum\limits_{<i,j>} 
{\hat{{\bf S}}}_i\cdot {\hat{{\bf S}}}_j 
-J'\sum\limits_{<i,j>}{\hat{S^z}}_i\,\,{\hat{S^z}}_j
-J_l\sum\limits_{i}{\hat{{\bf S}}}_i\cdot{\hat{{\bf s}}}_i-
\nonumber  \\
& & t \sum\limits_{<i,j>,\sigma} 
\left(\hat{c}^+_{i\sigma} \hat{c}_{j\sigma} + {\rm h.c.}\right)+ 
U\sum\limits_{i} \hat{n}_{i\uparrow} \hat{n}_{i\downarrow} - 
\mu\sum\limits_i \hat n_i+ \nonumber \\
& &  F\sum\limits_{<i,j>}\left(\hat{c}^+_{i\uparrow} \hat{c}^+_{i\downarrow}
\hat{c}_{j\downarrow} \hat{c}_{j\uparrow}+{\rm h.c.}\right)
\label{sfm1} 
\end{eqnarray} 
Here $\hat{c}^+_{i\sigma}$ and $\hat{c}_{i\sigma}$ are creation 
and annihilation operators for itinerant electrons, 
$\hat{n}_{i \sigma} = \hat{c}^+_{i \sigma}\hat{c}^{\phantom +}_{i \sigma}$ 
are density operators, 
${\hat{{\bf s}}}_{i}=1/2 \sum\limits_{\sigma\sigma'} \hat{c}^+_{i\sigma} 
{\vec {\tau}}_{\sigma\sigma'} \hat{c}^{\phantom +}_{i\sigma'}$, 
where ${\vec {\tau}}$ denotes the vector of Pauli matrices, are the spin operators
of itinerant electrons, and ${\hat{{\bf S}}}_i$ are spin-s operators of localized spins. 
The sums are over all sites of a three-dimensional lattice, 
$<i,j>$ denotes the sum over the nearest neighbors, and $\mu$ is the chemical 
potential. In (\ref{sfm1}) the $J$-term corresponds to a direct Heisenberg exchange of
localized spins which is ferromagnetic ($J>0$). The magnitude of the magnetocrystalline 
anisotropy is given by $J'$. Here I focus on uniaxial anisotropy, $J'>0$, with the
easy axis of magnetization along the $z$ axis. The local spin-fermion interaction is
ferromagnetic, too ($J_l>0$), and the $F$-term describes the hopping of local pairs 
consisting of spin-up and spin-down electrons ($F>0$)\cite{sfm11}. 

Still, the question remains whether the off-diagonal hopping parameters in the Hamiltonian,
which involve orbital overlaps between neighbouring sites, would be sufficiently large in 
view of the fact that $U-f$ orbitals are well localized. One expects that hybridization 
between $Ge$ electrons and $U-f$ electrons which gives an itinerant character to the 
f-electrons, leads to larger overlaps than pure $f$ orbitals.
 
I introduce Schwinger 
representation for the localized spin operators
${\hat{{\bf S}}}_i=1/2\sum\limits_{\sigma\sigma'}\hat f_{i,\sigma}^+
{\bf {\tau}}_{\sigma\sigma'}\hat f_{i,\sigma'}$,
where the bose operators satisfy 
the condition $\hat f_{i,\sigma}^{\dagger}\hat f_{i,\sigma}=2s$.
The partition function can be written as a path integral over the complex 
functions of the Matsubara time $\tau$,\,\, $f_{i\sigma}(\tau),\, 
f^+_{i\sigma}(\tau)$ and Grassmann functions 
$c^+_{i\sigma}(\tau)$ and $c_{i\sigma}(\tau)$ replacing the operators in 
the Hamiltonian Eqs.(\ref{sfm1}) with the functions\cite{negel}. In terms of Schwinger 
bosons the theory is $U(1)$ gauge invariant, where the bose fields have a charge 1, 
with respect to gauge transformations, while the fermi fields are gauge invariants. 

It is convenient to introduce two spin-singlet fermi fields
\begin{eqnarray}
\Psi^A_i(\tau) & = & \frac {1}{\sqrt {2s}}\left[f_{i1}(\tau)c_{i2}(\tau)\,-
\,f_{i2}(\tau)c_{i1}(\tau)\right] \nonumber \\
\Psi^B_i(\tau) & = & \frac {1}{\sqrt {2s}}f^+_{i\sigma}(\tau)c_{i\sigma}(\tau)
\label{sfm2} 
\end{eqnarray}
which are gauge variant with charge 1 and -1 with respect to gauge transformations.
Equations (\ref{sfm2})
can be regarded as a SU(2) transformation\cite{sfm12} and the Fermi measure is 
invariant under the change of variables. An important advantage is the fact that in terms 
of the spin-singlet Fermi fields the spin-fermion
interaction is diagonalized
$\sum\limits_{i}{{\bf S}}_i\cdot{{\bf s}}_i=s/2\sum\limits_{i}
[\Psi^{B+}_i\Psi^B_i-\Psi^{A+}_i\Psi^A_i]$
and one accounts for it exactly. The total spin of
the system ${\bf S}^{tot}_i={\bf S}_i+{\bf s}_i$ can be rewritten in the form
\begin{eqnarray}
{\bf S}_{tot}^i & = &  \frac 1s \left[s+\frac 12 
\left (\Psi^{B+}_i\Psi^B_i-\Psi^{A+}_i\Psi^A_i\right)\right]{\bf S}_i\,+ \nonumber \\
& & \frac 12 \Psi^{A+}_i\Psi^B_i {\bf T}_i\,+\,\frac 12 \Psi^{B+}_i\Psi^A_i{\bf T}^+_i
\label{sfm3}
\end{eqnarray}
where ${\bf S}_i$ is the spin vector of localized spins (${\bf S}^2_i=s^2$),
and ${\bf T}_i$ and ${\bf T}^+_i$ are complex vectors which depend on  Schwinger's 
bosons. They are orthogonal to the spin vector
${\bf S}_i\cdot{\bf T}_i={\bf S}_i\cdot{\bf T}^+_i=0$ and satisfy 
${\bf T}^2_i={\bf T}^{+2}_i=0, {\bf T}_i\cdot{\bf T}^+_i=2$. The gauge
invariance imposes the conditions $<\Psi^{A+}_i\Psi^B_i>=
<\Psi^{B+}_i\Psi^A_i>=0$. As a result, the dimensionless magnetization per
lattice site $M=<(S^{tot}_i)^z>$ reads 
\begin{equation}
M=\frac 1s \left[s+\frac 12 
<\left (\Psi^{B+}_i\Psi^B_i-\Psi^{A+}_i\Psi^A_i\right)>\right]<{\bf S}^{z}_i>
\label{sfm4}
\end{equation}
At zero temperature $<{\bf S}^{z}_i>=s$ and  
$M=s+m$, where
$m=1/2<\left (\Psi^{B+}_i\Psi^B_i-\Psi^{A+}_i\Psi^A_i\right)>$
is the contribution of the itinerant electrons.

Rewriting the Hamiltonian in terms of A and B fields, one obtains the following
representations for Hubbard and pair-hopping terms 
\begin{eqnarray}
& & \sum\limits_{i} n_{i\uparrow} n_{i\downarrow}=    
-\frac 12 \sum\limits_{i}\left (\Psi^{B+}_i\Psi^B_i-\Psi^{A+}_i\Psi^A_i\right)^2,
 \nonumber \\
& & \sum\limits_{<i,j>} c^+_{i\uparrow} c^+_{i\downarrow}
c_{j\downarrow} c_{j\uparrow}=
\sum\limits_{<i,j>}\Psi^{B+}_i\Psi^B_j\Psi^{A+}_i\Psi^A_j.
\label{sfm5}
\end{eqnarray}
One can decouple these terms by means of the Hubbard-Stratanovich 
transformation, 
introducing a real field $m_i(\tau)$ associated with the composite field
$1/2 (\Psi^{B+}_i\Psi^B_i-\Psi^{A+}_i\Psi^A_i)$, and complex fields $u^R_{ij}(\tau)$
associated with $\Psi^{R+}_i\Psi^R_j$, where $R$ stands for $A$ or $B$. Then, 
the action is quadratic with respect to the fermions and one can integrate them out.
The obtained free energy is a function 
of the composite fields and the integral over them can be 
performed approximately by means of the steepest descend method. To this end 
one sets the first derivatives of the free energy with respect to composite 
fields equal to zero. These are the mean-field equations. The solutions of the 
mean-field equations are assumed to be constants independent of the lattice sites and
bonds $m^0_i(\tau)=m, u^R_{ij}(\tau)=u^R$, where $m$ is the itinerant electron 
contribution to the magnetization(see Eq.(\ref{sfm4})).The equations for $m,u$ and
the number of itinerant electrons $n$ are
\begin{eqnarray}
m & = & \frac 12 \int\limits^{\frac D2}_{-\frac D2}d\epsilon N(\epsilon)
\left(f\left[\epsilon^B(\epsilon)\right]-f\left[\epsilon^A(\epsilon)\right]\right) 
\nonumber \\ 
\label{sfm6}
u^R & = & -\frac 2D \int\limits^{\frac D2}_{-\frac D2}d\epsilon N(\epsilon)\epsilon
f\left[\epsilon^R(\epsilon)\right],\quad R=A,B \\ 
n & = & \int\limits^{\frac D2}_{-\frac D2}d\epsilon N(\epsilon)
\left(f\left[\epsilon^B(\epsilon)\right]+f\left[\epsilon^A(\epsilon)\right]\right)
\nonumber
\end{eqnarray}
where $f\left[\epsilon^R(\epsilon)\right]$ is the Fermi function, $N(\epsilon)$ is the density
of state for band energy $\epsilon_k=-t\sum\limits_{\delta}e^{ik\delta}$ and bandwidth
$D=2zt$ with $\delta$ a vector connecting a site to its nearest neighbors and $z$
the number of nearest neighbors. In equations (\ref{sfm6}) the fermion dispersions are
\begin{eqnarray}
\epsilon^A(\epsilon_k) & = & \left(1-\frac Ft u^B\right)\epsilon_k+2mU+\frac {sJ_l}{2}-\mu,
\nonumber \\
\epsilon^B(\epsilon_k) & = & \left(1-\frac Ft u^A\right)\epsilon_k-2mU-\frac {sJ_l}{2}-\mu.
\label{sfm7}
\end{eqnarray}
I assume for simplicity a flat density of states: $N(\epsilon)=1/D,\,\,\,-D/2<\epsilon<D/2$.
Unlike in the Stoner model, the model with pair-hopping term does not depend strongly on
energy variation of the density of states\cite{sfm11}. Now the system can be analytically
solved at zero temperature. 

A solution with $m=0$ exists if $u^A$ and $u^B$ are nonzero and have opposite signs, which
in turn requires  $g=F/t>4$.
Then the equation for the contribution of itinerant electrons to the magnetization $m$ is
\begin{equation}
m^3+\left(\frac {2U}{Dg}-\frac {(n-1)^2}{4}-\frac 14\right)m - 
\left(\frac {|n-1|}{2g}-\frac {sJ_l}{2gD}\right)=0.
\label{sfm8}
\end{equation}

The equation (\ref{sfm8}) has a solution $m=0$ if $D=D_x$, where $D_x=sJ_l/|n-1|$.
The Coulomb parameter $U$ is large parameter in the theory, so one can choose it to 
satisfy $2U/Dg>(n-1)^2/4+1/4$. Then, the equation (\ref{sfm8}) has only one real solution. 

To match the experimental results it is most adequate to keep the parameters of the
local interactions $U$ and $J_l$, and the number of the itinerant electrons $n$ fixed.
I assume that hydrostatic pressure increases the pair-hopping at the expense of the single-electron
hopping. This means, that the 
pair-hopping parameter $F$ increases, while the hopping parameter $t$ decreases when the pressure increases.
At pressure $p=p_x$ $D=D_x$ and $F=F_x$, where $F_x=4U/z((n-1)^2+1)$. The first condition is necessary to have
a zero $m$ solution, the second one ensures an abrupt
decrease of magnetization at $p_x$. 
The last assumption is that when the pressure increases 
the parameter $F$ scales like $1/D$, more exactly $F/F_x=D_x/D$. Above $p_x$ the parameters of the 
itinerant electrons remain unchanged. One can find justification of this assumption in the experimental 
fact that above $p_x$ the physics of the system is dominated by the localized spins.   
It is important to 
stress that the transition, to the paramagnetism $M=0$, undergoes with
jump, because above $p_x$ the magnetization results from the localized spins.
The contribution of the itinerant electrons to the magnetization $m$ as a function of the 
pair-hopping parameter $F/F_x$ is depicted in fig.2 for $(n-1)^2=0.2$, and $1.6U=3sJ_l$.

\begin{figure}[h]  
\vspace{0.1cm}  
\epsfxsize=3.7cm  
\hspace*{-2.75cm}  
\epsfbox{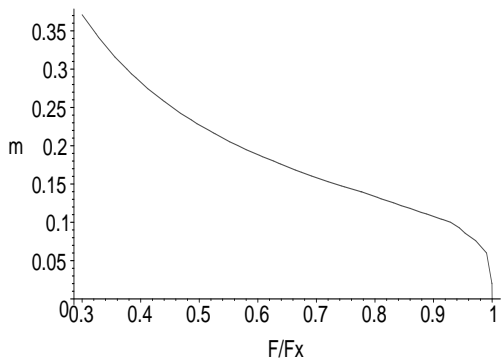}  
\caption{The contribution of the itinerant electrons to the magnetization $m$ as a function of the 
pair-hopping parameter $F/F_x$.}  
\label{fig2}  
\end{figure}  

The graph (fig.2) as well as the expressions for $F_x$ and $D_x$ are an artifact of the 
approximate treatment of the density of states $N(\epsilon)$. More accurate account
for the energy dependence of  $N(\epsilon)$ will give us different conditions for the
parameters and more realistic dependence of the magnetization on the parameters.   

The proposed model of $UGe_2$ differs from the models discussed in
\cite{sfm7,sfm8,sfm9,sfm10} in many aspects. First, degrees of freedom associated
with localized spins and itinerant electrons are introduced, which enables one 
to describe two different ferromagnetic phases $FM1$ and $FM2$ fig.1.
The resistivity measurements reveal\cite{sfm2} the presence of an additional
phase line that lies entirely within the ferromagnetic phase. It is suggested by
a strong anomaly seen in the resistivity\cite{sfm2,sfm13}. The characteristic temperature of
this transition, $T_x(p)$, decreases with pressure and disappears at a pressure
$p_x$ (IP) at which the superconductivity is strongest. For pressures below $p_x$
the itinerant electrons contribute to the magnetization, while for pressures above $p_x$
the ferromagnetism is dominated by localized spins. This suggests 
to define $T_x$ by the equation $m(T_x)=0$. 
Above $T_x$ the itinerant electrons do not contribute to 
the magnetization, and the ferromagnetism is entirely dominated by the spin fluctuations
of the localized spins, while below $T_x$ the itinerant electrons take part in 
the formation of the spin fluctuations. 
In particular, the itinerant electron mass renormalization is 
different below $T_x$ and above this temperature. As a result, the slope in the 
$d\rho/dT$ versus $T$ diagram is different above $T_x$ and well below $T_x$. 
Increasing the temperature from below $T_x$ the slope changes smoothly
from its value well below $T_x$ to its value above $T_x$, This means  
a non-Fermi-liquid temperature dependence of the resistivity within a temperature
interval around $T_x$\cite{sfm13}. The present description of the ferromagnetism above and below $T_x$
is in very good agreement with the experimental finding that the high pressure ferromagnetic
phase might have the more localized character\cite{sfm6}.

Second, the model explains in a unified way the superconductivity and
$T_x$ transition near $p_x$ point. At $p_x$ the contribution of the itinerant electron to the
magnetization $m$ becomes equal to zero and hence it is the end of the $T_x$ line, as follows
from the definition above. On the other side, it was explained that when $m=0$ the 
superconducting critical temperature is highest.

This work was supported by the Sofia University Grant 643/2002

\end{document}